\documentclass[11pt]{article}
\usepackage{moriond,epsfig}
\graphicspath{{ps/}}

\def\be{\begin{equation}}
\def\ee{\end{equation}}
\def\bea{\begin{eqnarray}}
\def\eea{\end{eqnarray}}

\begin{document}
\vspace*{4cm}
\title{\bf \boldmath MEASUREMENTS OF $\phi_1$ AND $\phi_2$ BY BELLE AND BABAR}

\author{T.A.-Kh.~Aushev \footnote{{\bf e-mail}: aushev@itep.ru}$^{\,1,2}$\\
for the Belle Collaboration\\
$^1$\small{\em Swiss Federal Institute of Technology of Lausanne, EPFL}\\
$^2$\small{\em Institute for Theoretical and Experimental Physics, Moscow, ITEP}\\
}

\address{Lausanne, Switzerland}

\maketitle\abstracts{
We report recent measurements of the Unitarity triangle angles
$\phi_1$ and $\phi_2$ using large data samples collected with Belle
and BaBar detectors at $e^+e^-$ asymmetric-energy colliders.
}

\section{Introduction}

In the Standard Model (SM), $CP$ violation in $B^0$ meson decays
originates from an irreducible complex phase in the $3\times3$
Cabibbo-Kobayashi-Maskawa (CKM) mixing matrix~\cite{ckm}.  The angles
$\phi_1$ and $\phi_2$ of the CKM unitarity triangle have been measured
in several $B$ decay
modes~\cite{jpsiks_belle,jpsiks_babar,phi2_belle,phi2_babar}.  Extra
studies in different decay modes are important to check the
self-consistence between measurements to probe the existence of New
Physics.

The results reported in this paper were obtained by two experiments,
Belle and BaBar, working at $e^+e^-$ asymmetric-energy colliders,
KEKB~\cite{KEKB} and PEP-II, correspondingly, with the center-of-mass
(CM) energy at $\Upsilon(4S)$ resonance ($\sqrt s=10.58\,{\rm GeV}$).
The Belle detector~\cite{belledet} is a large-solid-angle magnetic
spectrometer that consists of a silicon vertex detector (SVD), a
50-layer central drift chamber (CDC), a mosaic of aerogel threshold
Cherenkov counters (ACC), time-of-flight scintillation counters (TOF),
and an array of CsI(Tl) crystals (ECL) located inside a
superconducting solenoid coil that provides a $1.5$~T magnetic field.
An iron flux-return located outside of the coil is instrumented to
detect $K_L$ mesons and to identify muons (KLM).  For the results from
Belle experiment the data sample of 657 million $B\bar B$ pairs is
used.

The BaBar detector is described in detail elsewhere~\cite{babardet}.
Charged particle momenta are measured with a tracking system consisting 
of a five-layer silicon vertex tracker (SVT) and a 40-layer drift chamber
(DCH) surrounded by a 1.5~T solenoidal magnet. An electromagnetic
calorimeter (EMC) comprising 6580 CsI(Tl) crystals is used to measure
photon energies and positions.  Charged hadrons are identified with a
detector of internally reflected Cherenkov light (DIRC) and ionization
measurements in the tracking detectors.  The results from BaBar
experiment are based on 383 million $B\bar B$ pairs data sample.

\section{Study of $B^+\to D^+\bar D^0$ and search for $B^0\to D^0\bar D^0$}

Recently, evidence of direct $CP$ violation in the decay $B^0\to
D^+D^-$ was observed by Belle~\cite{dpdm_belle}, while BaBar measured
an asymmetry consistent with zero~\cite{dpdm_babar}.  A similar effect
might occur in the charged mode $B^+\to D^+\bar D^0$~\cite{cc}.  This
decay has already been observed by Belle~\cite{dpd0_belle} and
confirmed by BaBar~\cite{dpd0_babar}.

\begin{figure}[htb]
\centering
\includegraphics[width=0.35\textwidth]{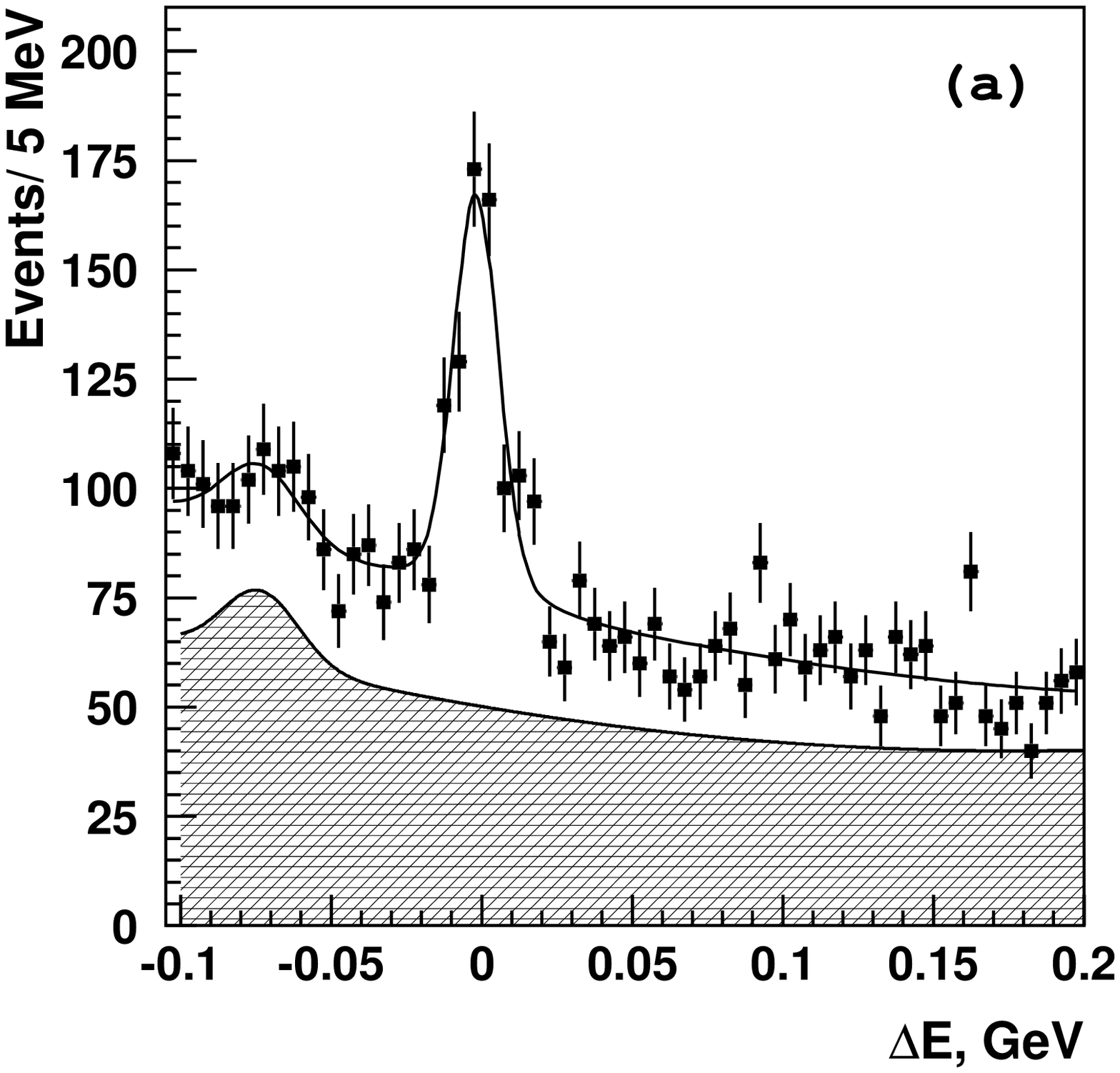}
\includegraphics[width=0.35\textwidth]{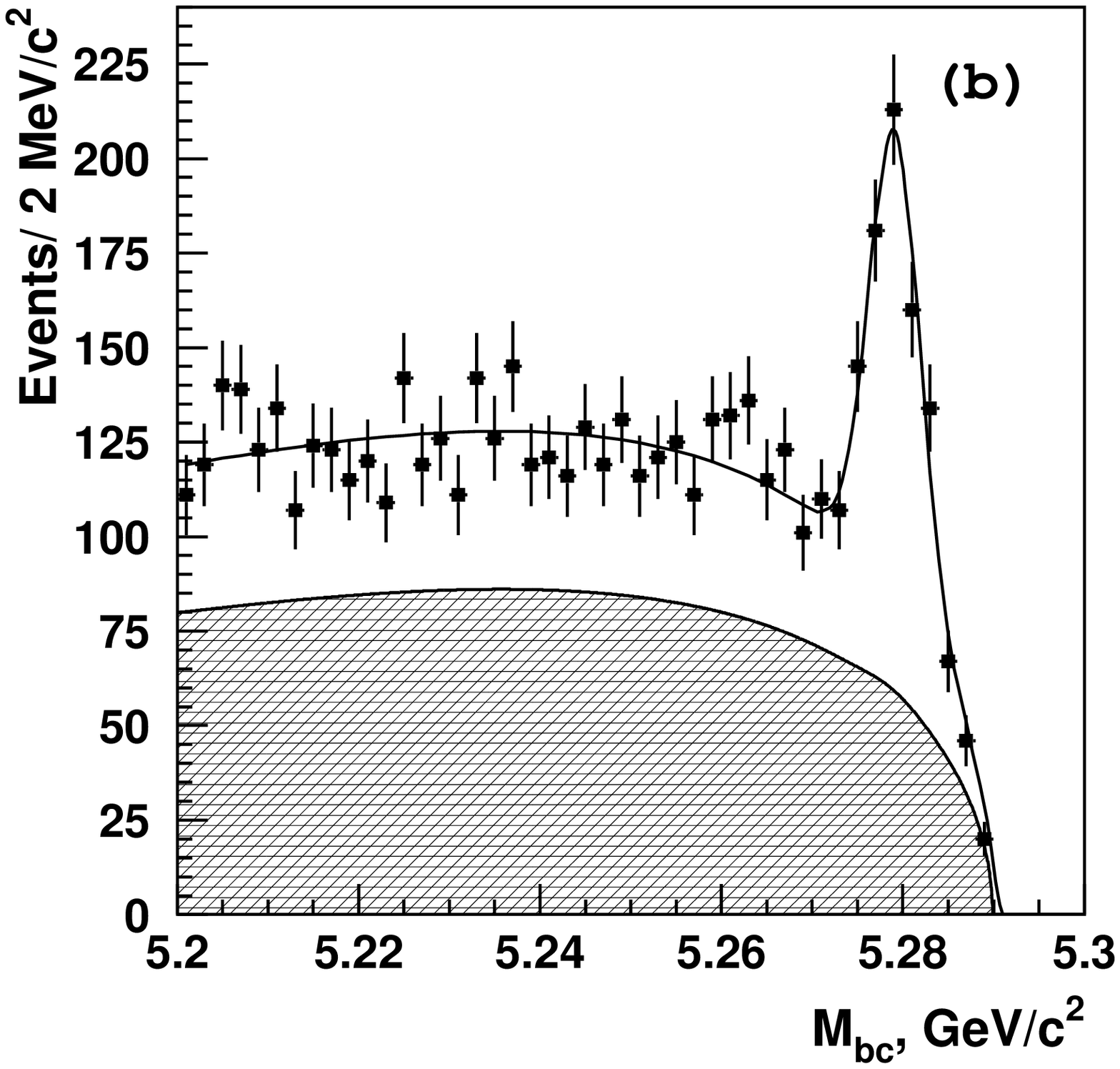}\\
\includegraphics[width=0.35\textwidth]{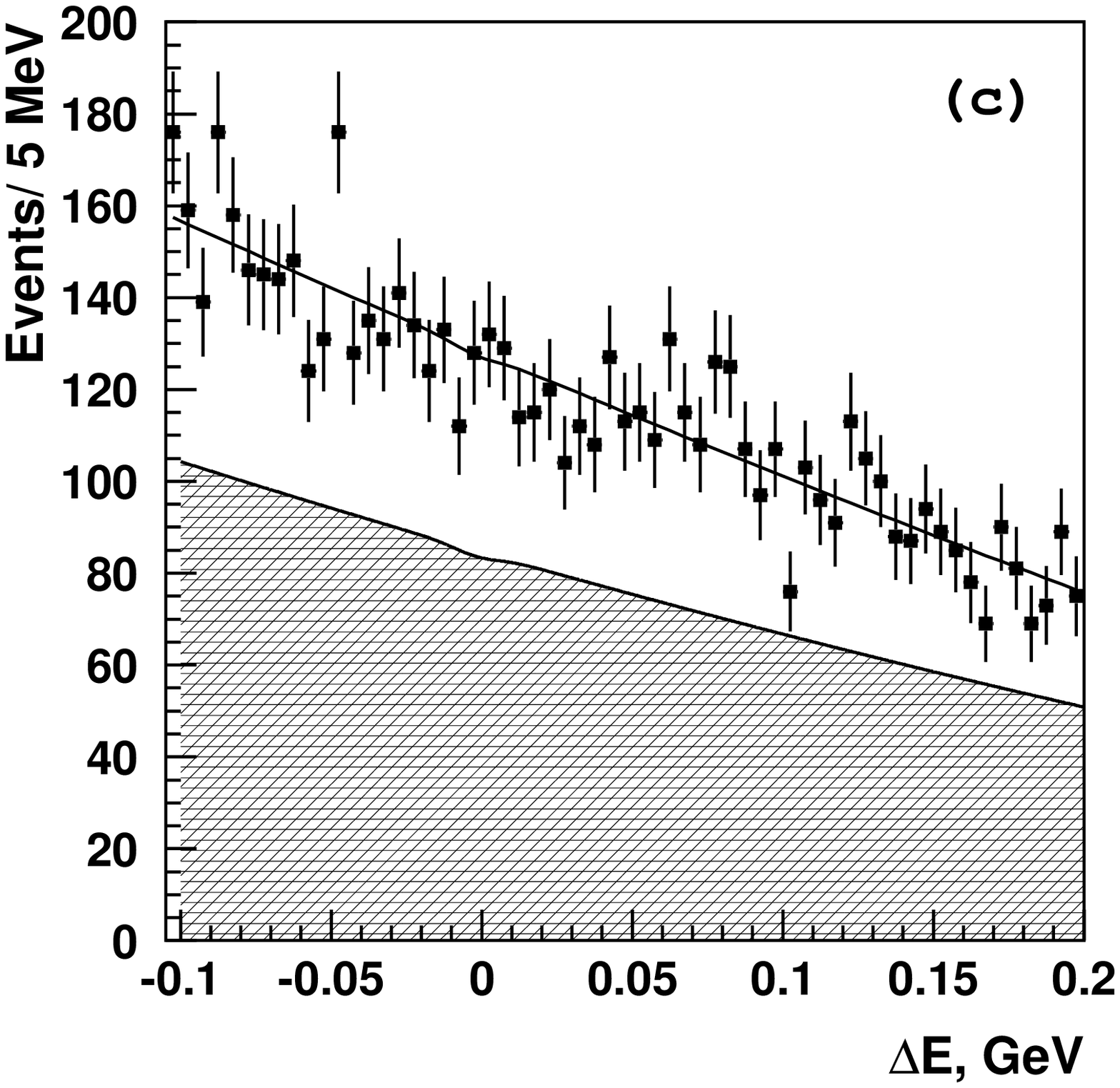}
\includegraphics[width=0.35\textwidth]{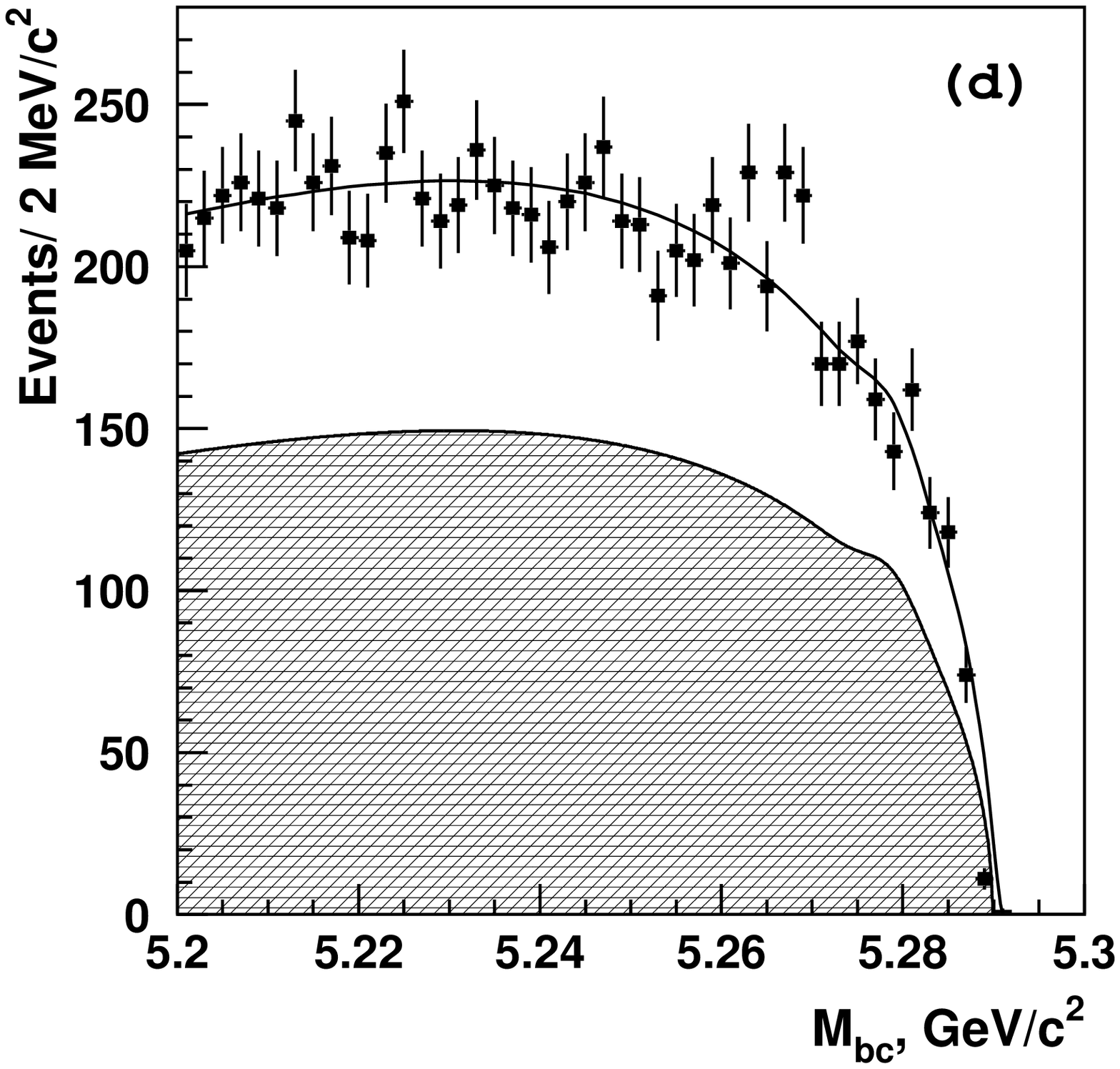}
\caption{$\Delta E$ (a,c) and $M_{\rm bc}$ (b,d) distributions for the
  $B^+\to D^+\bar D^0$ (a,b) and $B^0\to D^0\bar D^0$ (c,d)
  candidates.  Each distribution is the projection of the signal
  region of the other parameter. Points with errors represent the
  experimental data, open curves show projections from the $2D$ fits
  and cross-hatched curves show the $B\bar B$ background component
  only.}
\label{dd}
\end{figure}

Now, Belle updated their result with larger data
sample~\cite{dd0_belle}.  $366\pm32$ events were found from the fit to
the $\Delta E-M_{\rm bc}$ distribution (Fig.~\ref{dd}(a,b)), where
$\Delta E=E_B-E_{\rm beam}$, $M_{\rm bc}=\sqrt{E_{\rm
    beam}^2-p_B^{*2}}$, $E_B (p_B^*)$ is the energy (momentum) of $B$
candidate in the CM system, $E_{\rm beam}$ is the CM beam energy.  The
branching fraction of $B^+\to D^+\bar D^0$ is measured to be ${\cal
  B}(B^+\to D^+\bar D^0)=(3.85\pm0.31\pm0.38)\times10^{-4}$, where the
first error is statistical and the second one is systematic.  The
charge asymmetry for this decay is measured to be consistent with
zero: $A_{CP}(B^+\to D^+\bar D^0)=0.00\pm0.08\pm0.02$.  Belle also
searched for the decay $B^0\to D^0\bar D^0$.  An upper limit is
established for the branching fraction: ${\cal B}(B^0\to D^0\bar
D^0)<0.43\times10^{-4}$ (Fig.~\ref{dd}(c,d)).

\section{Study of $B^0\to D^{*+}D^{*-}$}

\begin{figure}[htb]
\centering
\includegraphics[width=0.6\textwidth]{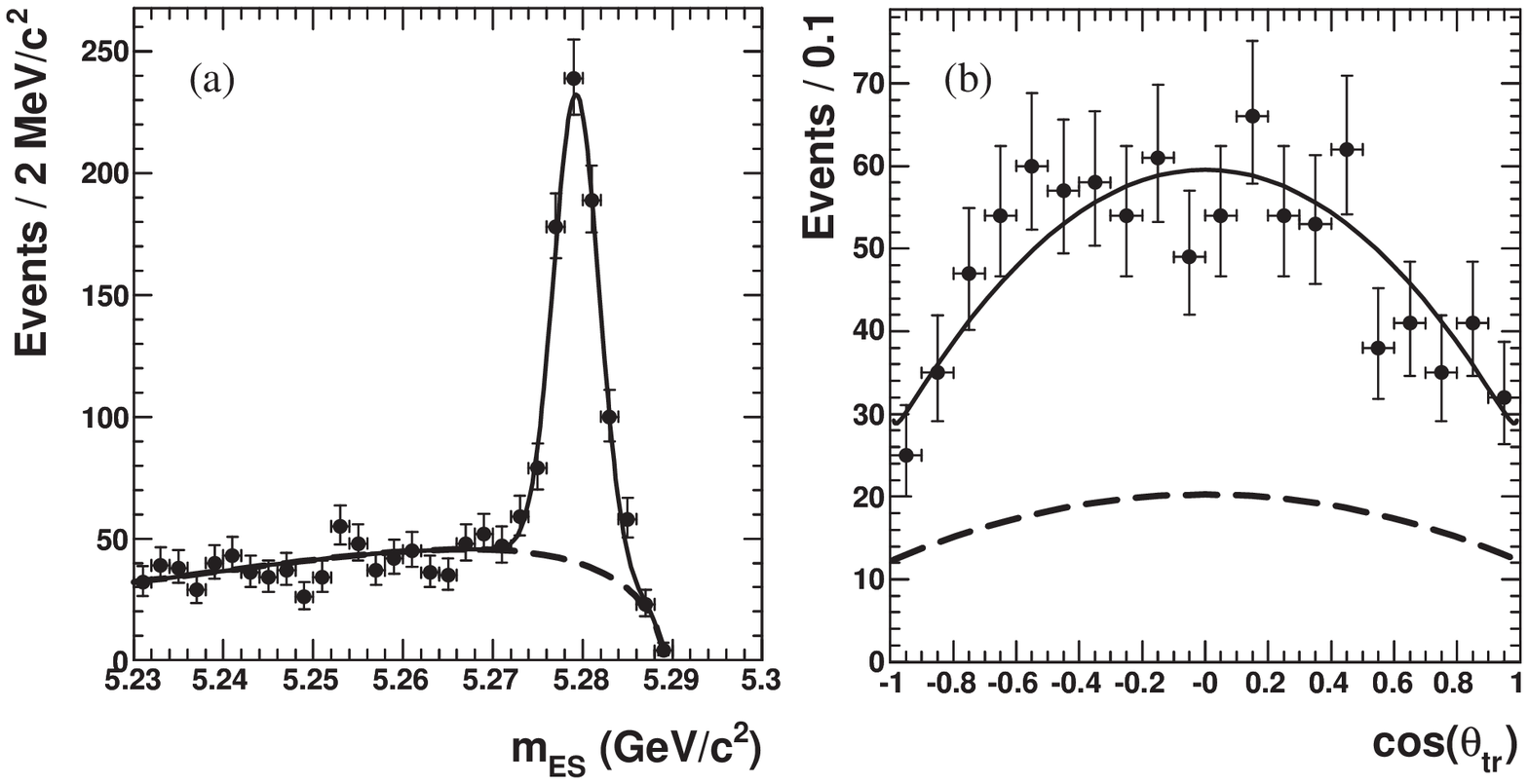}\\
\includegraphics[width=0.3\textwidth]{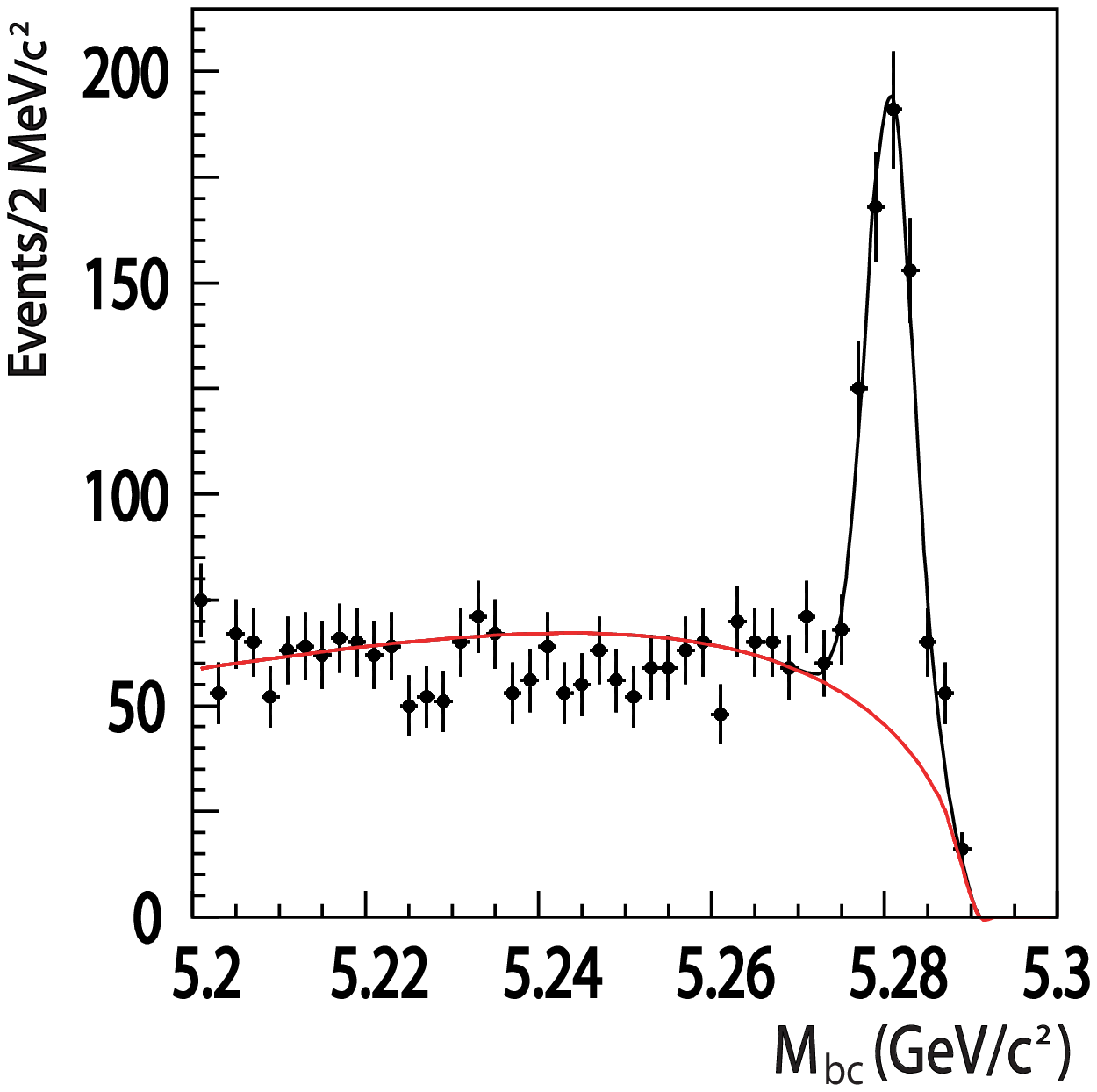}
\includegraphics[width=0.3\textwidth,height=0.2\textheight]{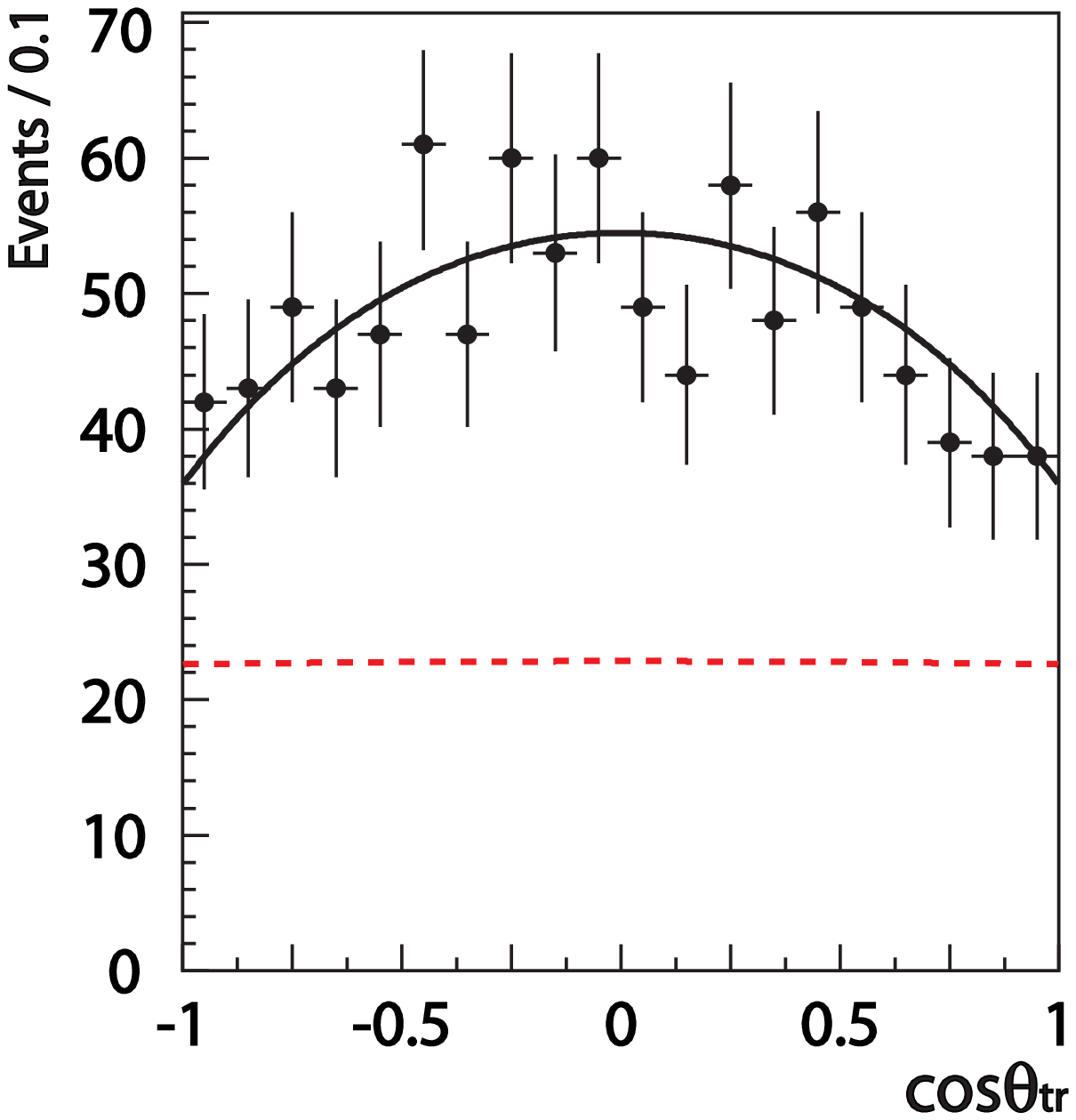}
\put(-240,120){\bf\tiny (c)}
\put(-105,120){\bf\tiny (d)}
\caption{Measured distributions of $M_{\rm bc}$ $(a,c)$ and
  $\cos\theta_{\rm tr}$ in the region $M_{\rm bc}>5.27$~GeV/$c^2$
  $(b,d)$ for BaBar $(a,b)$ and Belle $(c,d)$ of $B^0\to D^{*+}D^{*-}$
  events.  The solid lines are the projections of the fit results.
  The dotted lines represent the background components.}
\label{2dst_signal}
\end{figure}

Another interesting decay mode to study the $CP$ asymmetry is $B^0\to
D^{*+}D^{*-}$.  Both experiments have updated their results for this
decay mode and obtained high statistics signals shown in
Fig.~\ref{2dst_signal}(a,c)~\cite{2dst_babar}.  The time-dependent
decay rates of $B^0$ and $\bar B^0$ to a $CP$ eigenstate, like
$D^{*+}D^{*-}$, is given by formula:
\[
{\cal P}(\Delta t)=
\frac{e^{-\Delta t/\tau_{B^0}}}{4\tau_{B^0}}
\Big\{1+q\Big[{\cal S}_{f_{CP}}\sin(\Delta m_d\Delta t_{B^0})+
  {\cal A}_{f_{CP}}\cos(\Delta m_d\Delta t_{B^0})\Big]\Big\},
\]
where $q$ is the $b$-flavor charge: $q=+1(-1)$ when the tagging $B$
meson is a $B^0$ ($\bar B^0$), $\tau_{B^0}$ is the neutral $B$
lifetime, $\Delta m_d$ is the mass difference between two $B^0$ mass
eigenstates, $\Delta t_{B^0}=t_{CP}-t_{\rm tag}$.  The tree diagram
dominates in this decay mode, which according to the SM gives ${\cal
  S}_{f_{CP}}=\xi_{D^{*+}D^{*-}}\sin2\phi_1$ and ${\cal
  A}_{f_{CP}}=0$.  The parameter $\xi_{D^{*+}D^{*-}}$ is the $CP$
eigenvalue of the $D^{*+}D^{*-}$, which is $+1$ when the decay
proceeds via $S$ and $D$ waves, or $-1$ for a $P$ wave.  Therefore,
the $CP$ measurement requires helicity study to obtain the $CP$-odd
fraction $R_{\rm odd}$ of the decay.  It is done in both analyses from
Belle and BaBar in the so-called transversity basis.  The fit results
are presented in Fig.~\ref{2dst_signal}(b,d).  The parameter $R_{\rm
  odd}$ is found to be equal to $0.143\pm0.034({\rm
  stat})\pm0.008({\rm syst})$ by BaBar and $0.116\pm0.042({\rm
  stat})\pm0.004({\rm syst})$ by Belle.

\begin{figure}[htb]
\centering
\includegraphics[width=0.35\textwidth]{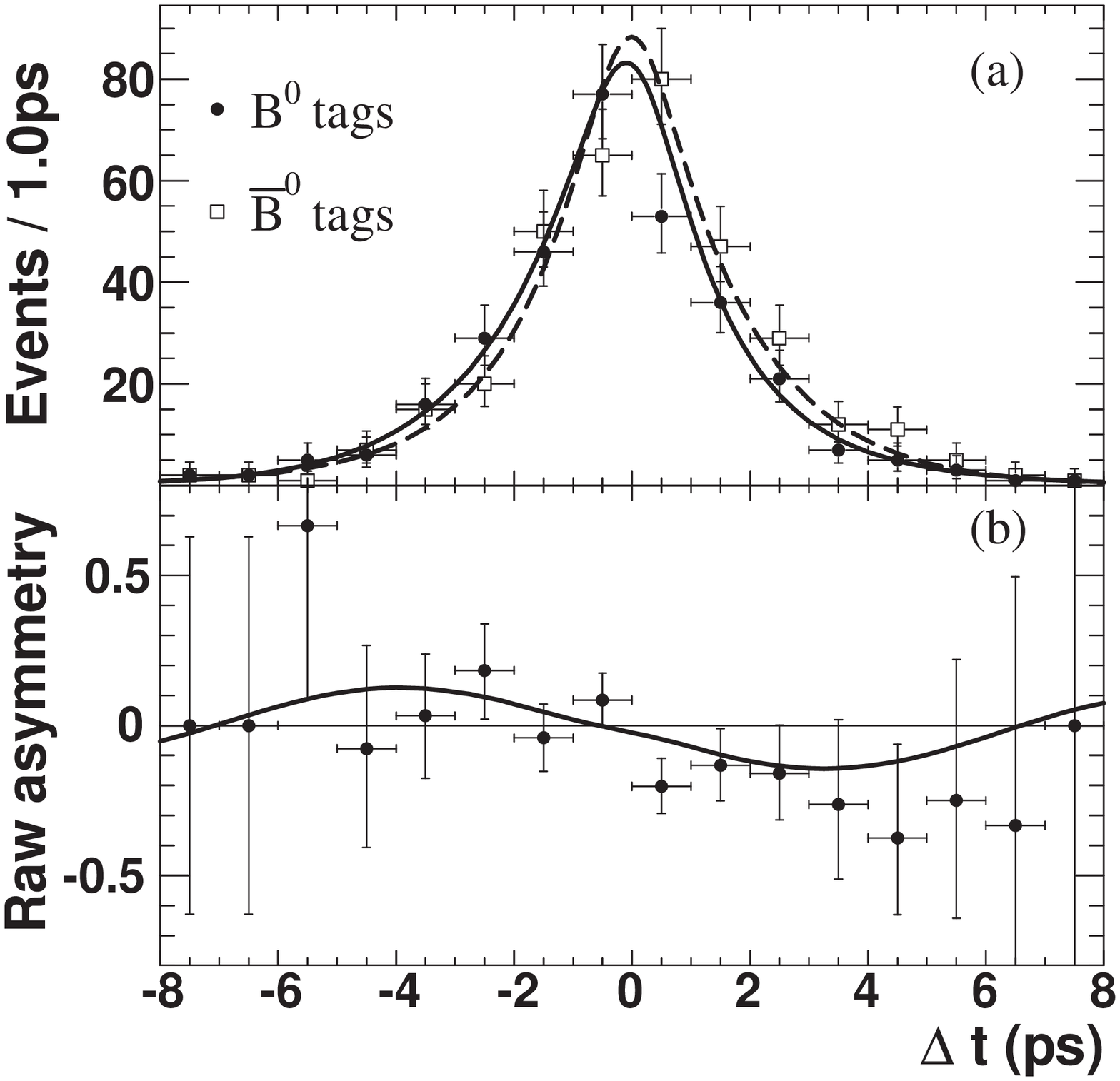}
\includegraphics[width=0.35\textwidth,height=.22\textheight]{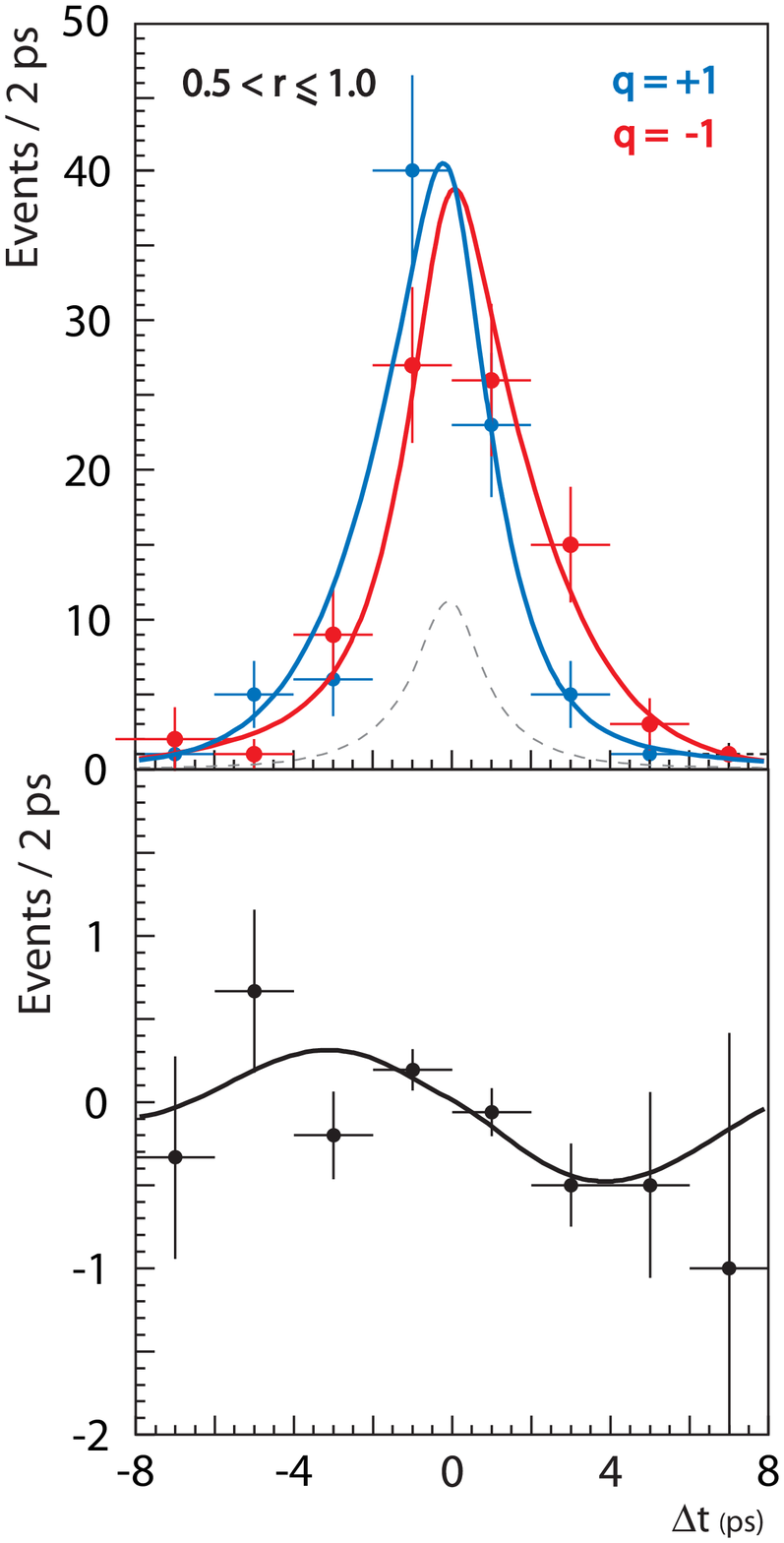}
\put(-30,125){\bf\tiny (c)}
\put(-30, 60){\bf\tiny (d)}
\caption{The $\Delta t$ distributions of $B^0\to D^{*+}D^{*-}$ events
  in the region $M_{\rm bc}>5.27$~GeV/$c^2$ for $B^0(\bar B^0)$ tagged
  candidates $(a,c)$ and the raw asymmetry $(N_{B^0}-N_{\bar
    B^0})/(N_{B^0}+N_{\bar B^0})$, as a function of $\Delta t$ $(b,d)$
  for BaBar $(a,b)$ and Belle $(c,d)$.  The lines represent the fit
  results.}
\label{2dst_cp}
\end{figure}

Finally, the unbinned maximum likelihood fit was performed to obtain
the $CP$-violating parameters.  The results of the fits are summarized
in Table~\ref{2dst_table} and presented in Fig.~\ref{2dst_cp}.  Both
experiments obtained the results well consistent with each other in
both the $CP$-odd fraction and the $CP$-violating parameters.  Note
that in the BaBar parametrization ${\cal A}=-{\cal C}$.  The Belle
results are preliminary.
\begin{table}
\caption{Results for $B^0\to D^{*+}D^{*-}$ decay mode.}
\centering
\begin{tabular}{|lcccc|}
\hline
      & Yield & $R_{\rm odd}$ & ${\cal A}=-{\cal C}$ & ${\cal S}$ \\
\hline
Belle & $545\pm29$ & $0.116\pm0.042\pm0.004$ & $+0.16\pm0.13\pm0.02$ & $-0.93\pm0.24\pm0.15$\\
BaBar & $638\pm38$ & $0.143\pm0.034\pm0.008$ & $+0.02\pm0.11\pm0.02$ & $-0.66\pm0.19\pm0.04$\\
\hline
\end{tabular}
\label{2dst_table}
\end{table}

\section{$CP$-violation in $B^0\to K_S\pi^0\pi^0$ and $B^0\to K_S\pi^0$}

In the SM, the $CP$ violation parameters in $b\to s$ ``penguin'' and
$b\to c$ ``tree'' transitions are predicted to be the same, ${\cal
  S}_f\approx-\xi_f\sin2\phi_1$ and ${\cal A}_f\approx0$, with small
theoretical uncertainties.  Recent measurements however, indicate that
the effective $\sin2\phi_1$ value, $\sin2\phi_1^{\rm eff}$, measured
with penguin processes is different from $\sin2\phi_1=0.687\pm0.025$
measured in tree decays by $2.6\sigma$~\cite{b2s_diff}.  New particles
in loop diagrams may shift the weak phase.

\begin{figure}[htb]
\includegraphics[width=0.25\textwidth,height=.24\textheight]{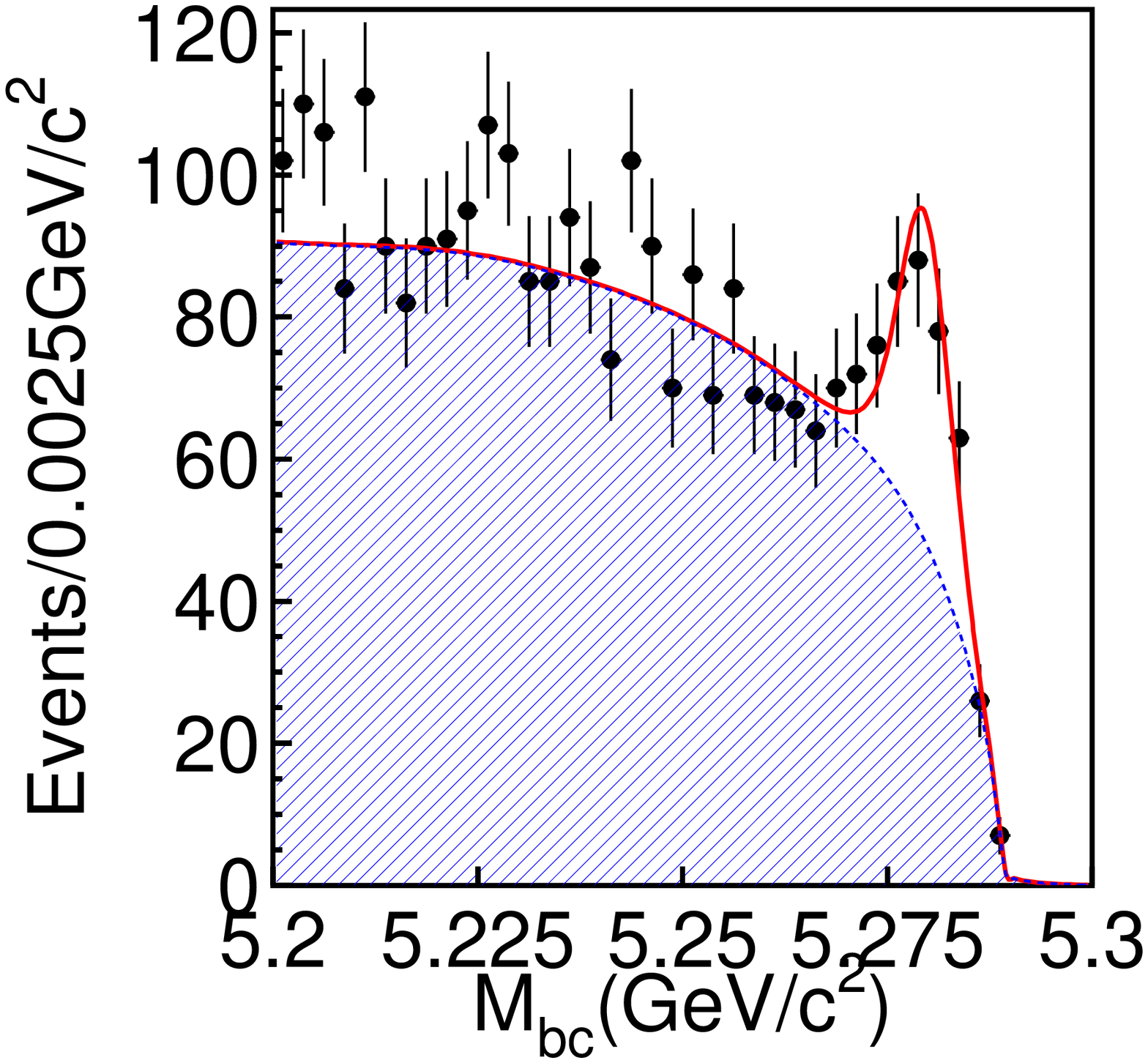} 
\put(-40,150){\bf\tiny (a)}
\includegraphics[width=0.25\textwidth,height=.27\textheight]{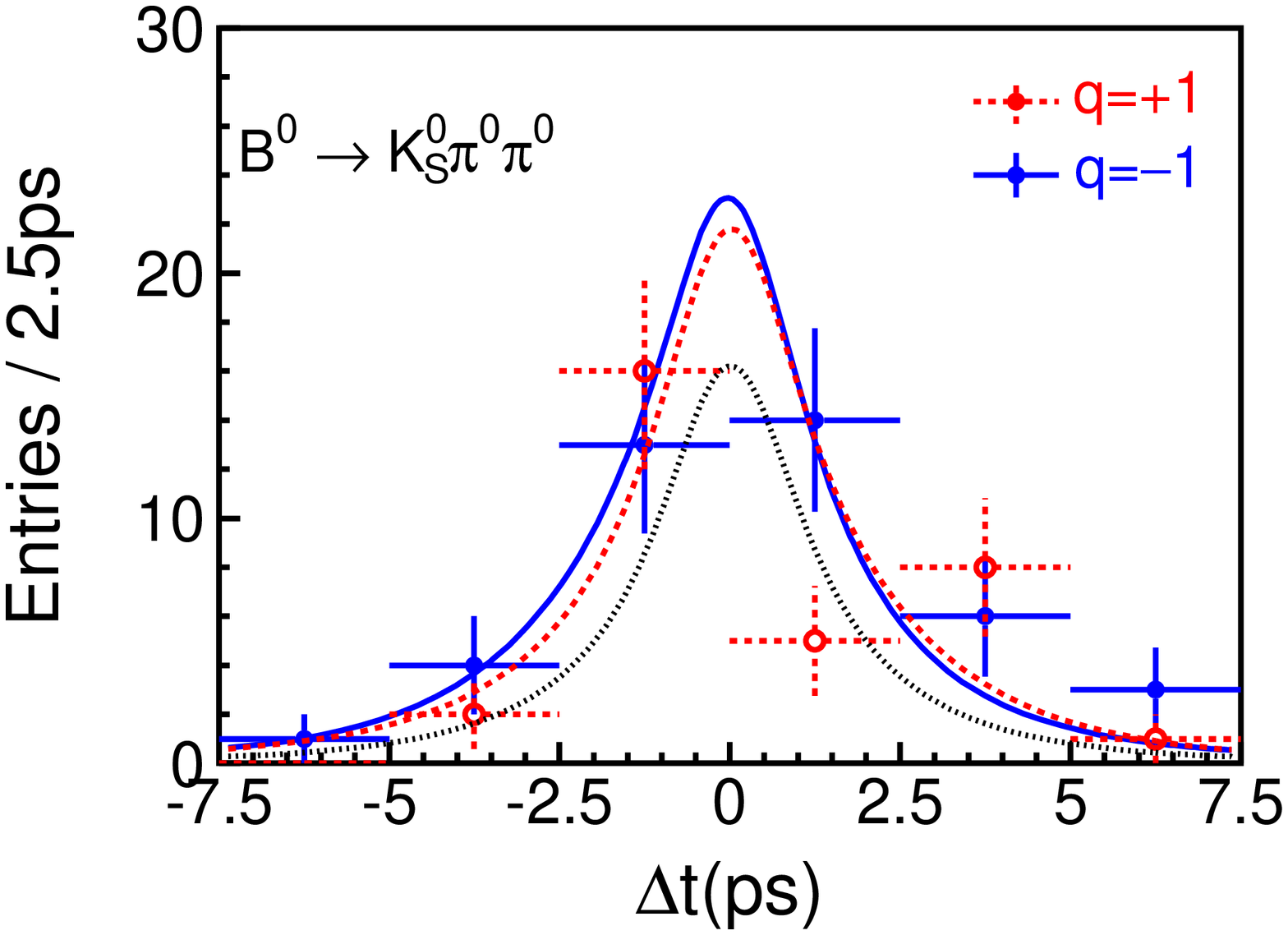} 
\put(-65,150){\bf\tiny (b)}
\includegraphics[width=0.25\textwidth,height=.27\textheight]{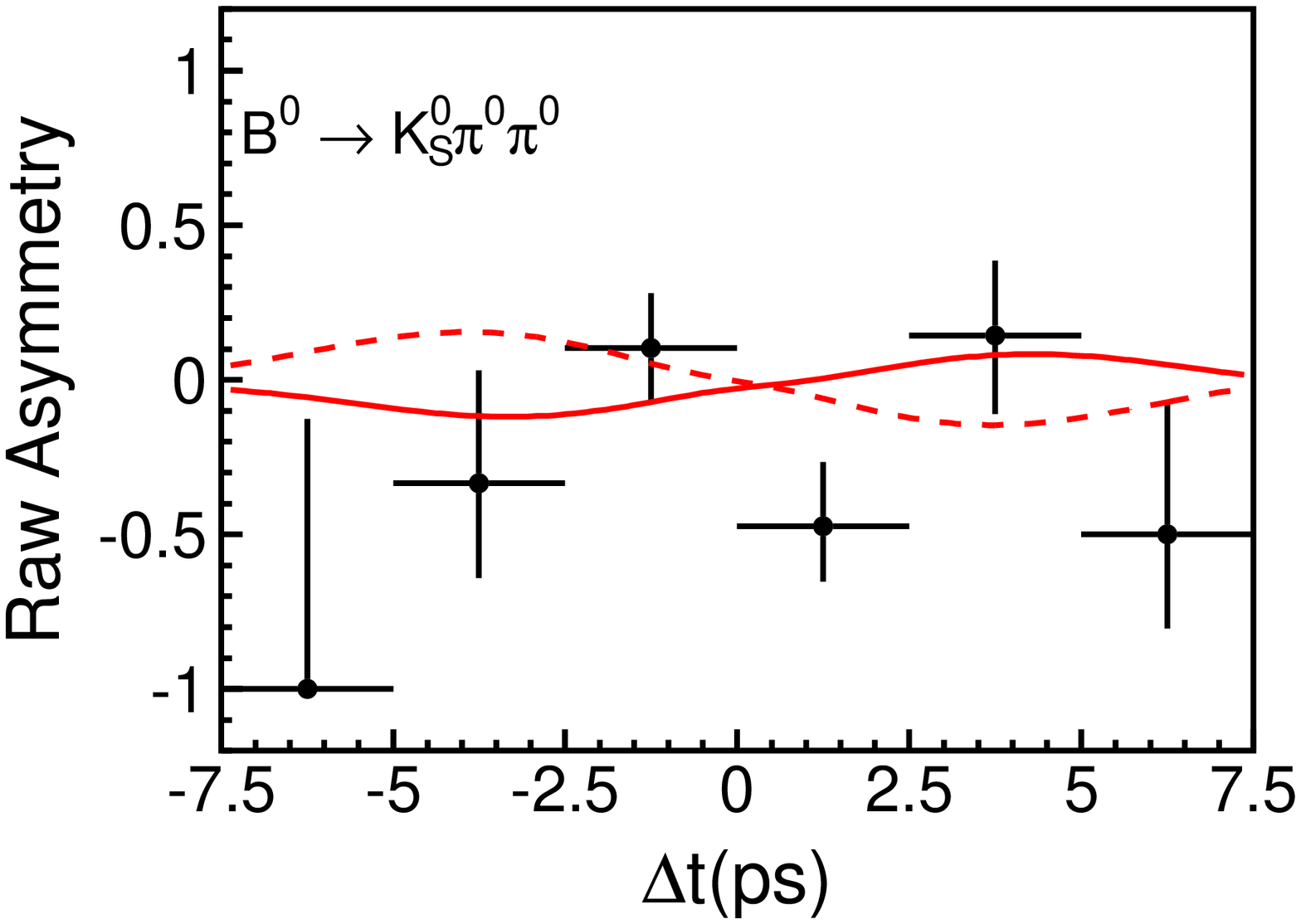}
\put(-40,150){\bf\tiny (c)}
\includegraphics[width=0.25\textwidth,height=.25\textheight]{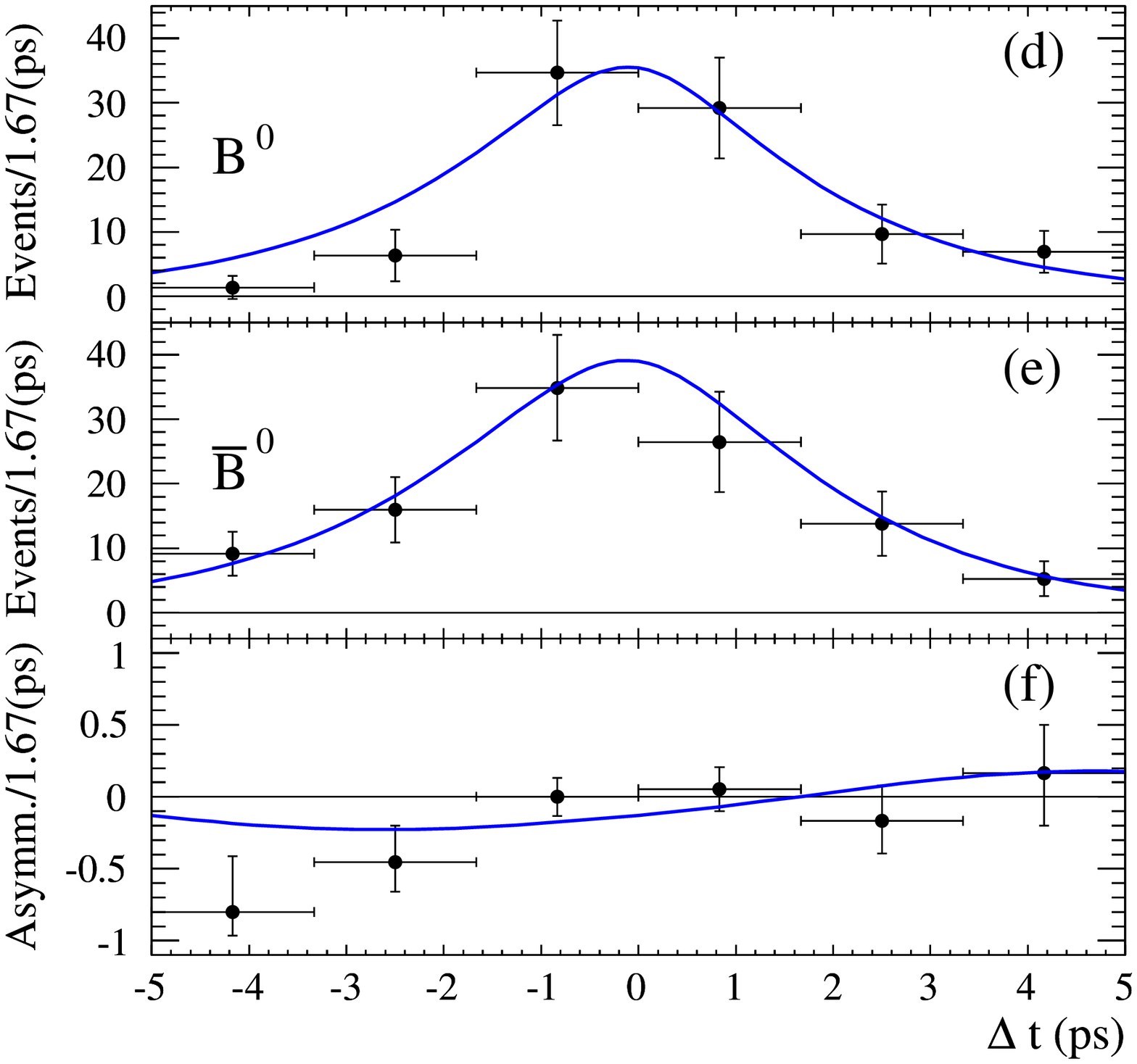}
\caption{Distributions for the $M_{\rm bc}$ (a), the $\Delta t$ (b)
  and the raw asymmetry (c) for $B^0\to K_S^0\pi^0\pi^0$ decay mode
  from Belle.  $\Delta t$ distributions for the $B^0$ (d) and $\bar
  B^0$ (e) tagged events and the raw asymmetry (f) for the decay
  $B^0\to K_S^0\pi^0$ from BaBar.  The lines represent the fit
  result.}
\label{kspi}
\end{figure}

Recently, Belle and BaBar measured the $CP$ asymmetry in $B^0\to
K_S^0\pi^0\pi^0$ and $B^0\to K_S\pi^0$ decays that proceed through
$b\to s\bar qq (q=u,d)$
transitions~\cite{kspi0pi0_belle,kspi0pi0_babar,kspi0_belle,kspi0_babar}.
The results of $CP$-violating parameters measurements are presented in
Table~\ref{kspi_table}.  Both experiments are perfectly consistent
with each other.  In the case of $B^0\to K_S^0\pi^0\pi^0$ the central
value of ${\cal S}$ has a sign opposite to what we expect from the SM,
but the errors are still too large to claim the contradiction.  The
estimated deviation of the average value from the SM is more than
$2\sigma$.  The fit to the data for Belle for $B^0\to K_S^0\pi^0\pi^0$
is presented in Fig.~\ref{kspi}(a-c) and the BaBar result for $B^0\to
K_S^0\pi^0$ is shown in Fig.~\ref{kspi}(d-f).
\begin{table}
\caption{Results for $B^0\to K_S^0\pi^0\pi^0$ and $B^0\to K_S^0\pi^0$ 
decay modes.}
\centering
\begin{tabular}{|lcc|}
\hline
        & ${\cal A}=-{\cal C}$ & ${\cal S}=-\sin2\phi_1$ \\
\hline
$B^0\to K_S^0\pi^0\pi^0$ & & \\
\hline
Belle   & $-0.17\pm0.24\pm0.06$ & $+0.43\pm0.49\pm0.09$\\
BaBar   & $-0.23\pm0.52\pm0.13$ & $+0.72\pm0.71\pm0.08$\\
Average & $-0.18\pm0.22$        & $+0.52\pm0.41$\\
\hline
$B^0\to K_S^0\pi^0$ & & \\
\hline
Belle   & $-0.05\pm0.14\pm0.05$ & $+0.33\pm0.35\pm0.08$\\
BaBar   & $-0.24\pm0.15\pm0.03$ & $+0.40\pm0.23\pm0.03$\\
\hline
\end{tabular}
\label{kspi_table}
\end{table}

\section{$\phi_2$ measurements}

\begin{figure}[htb]
\begin{tabular}{cc}
\includegraphics[width=0.33\textwidth]{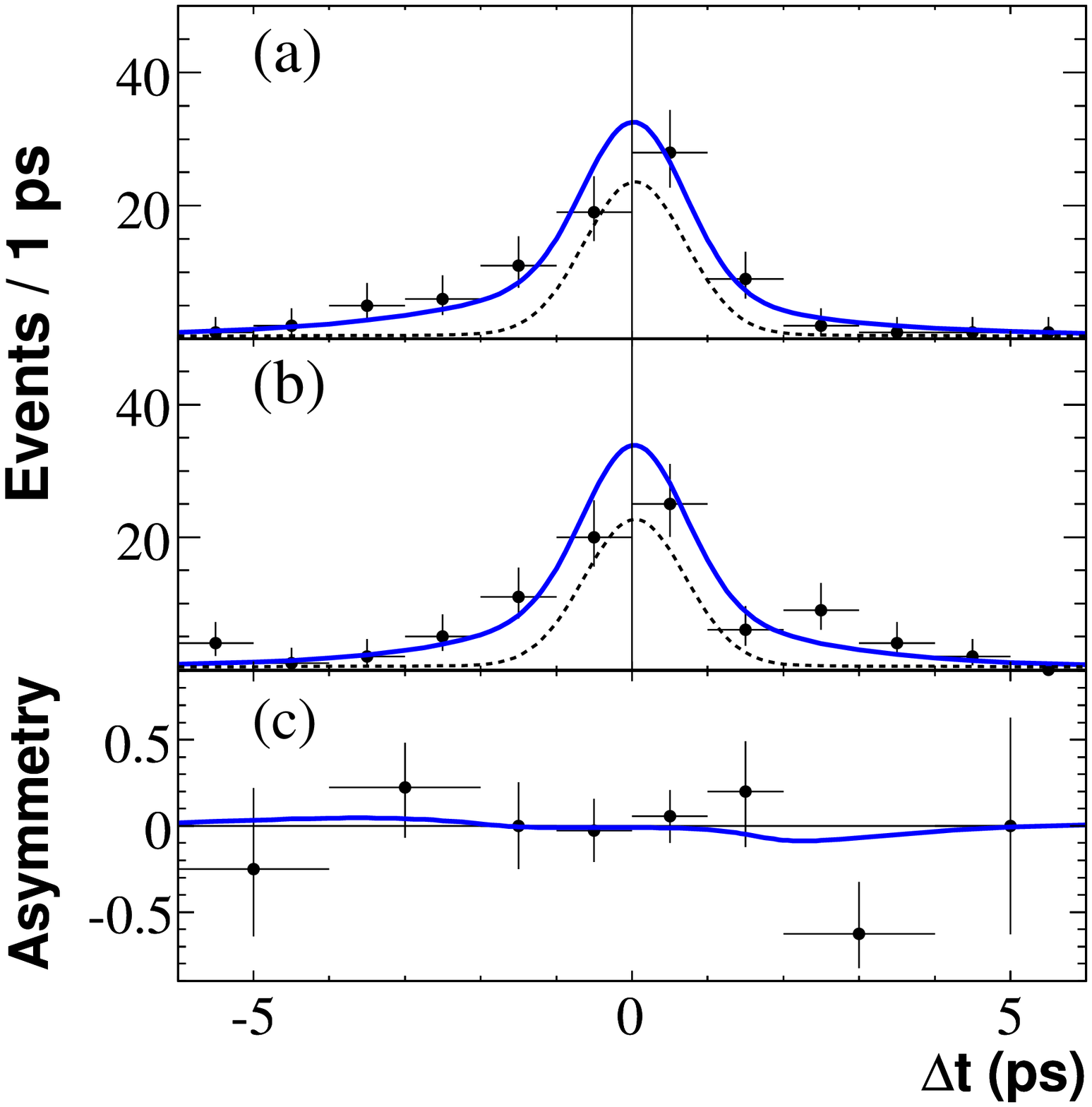}
\includegraphics[width=0.33\textwidth]{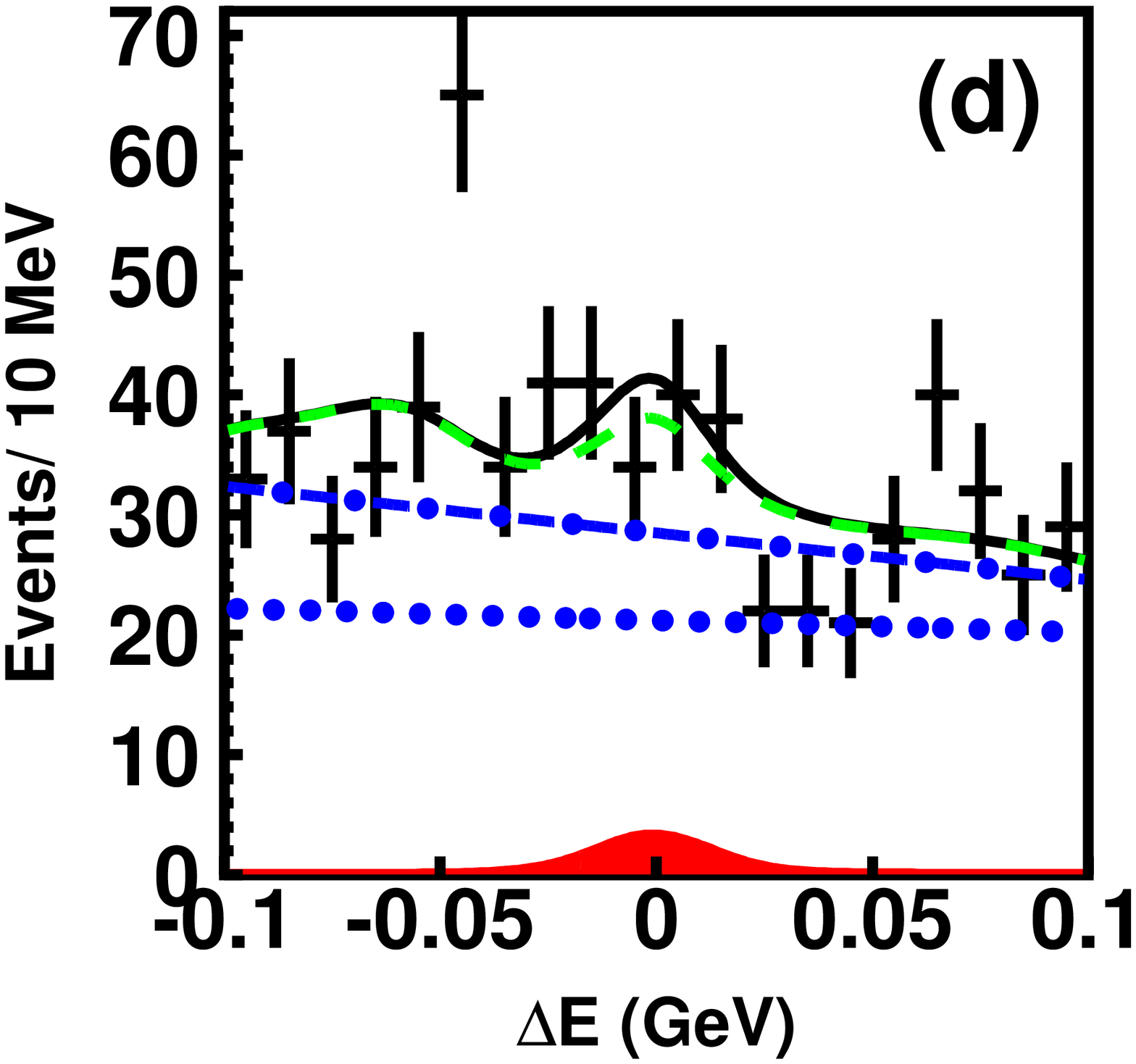}
\includegraphics[width=0.33\textwidth]{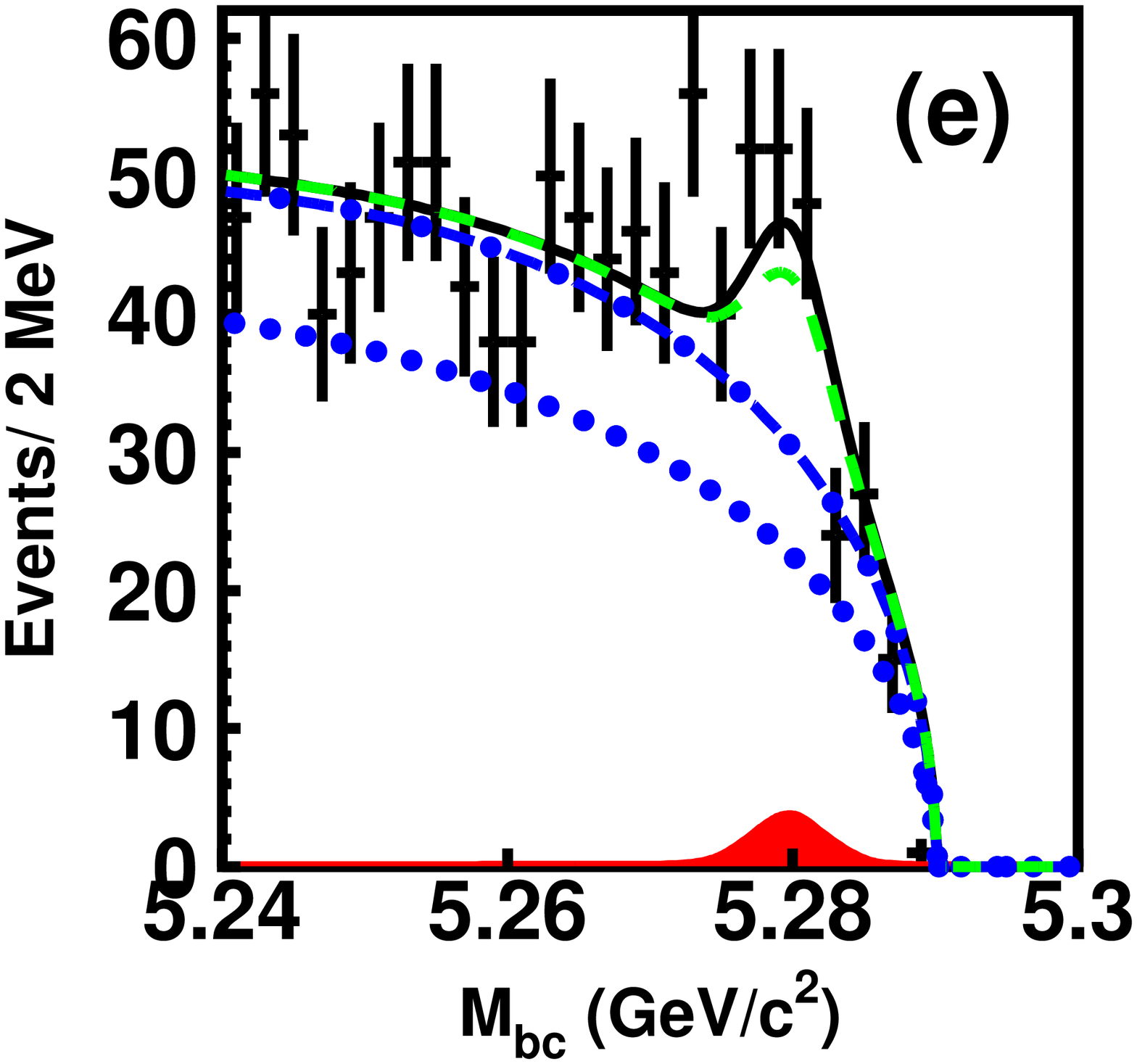}
\end{tabular}
\caption{$\Delta t$ distributions of the decay $B^0\to a_1^\pm\pi^\mp$
  for $B^0$ (a) and $\bar B^0$ (b) tagged events, and the raw
  asymmetry (c).  The solid lines show the fit results, while the
  dotted lines show the background component.  Projection of the
  signal region onto (d) $\Delta E$ and (e) $M_{\rm bc}$ for
  $B^0\to\rho^0\rho^0$ candidates.  The fit result is shown as the
  thick solid curve; the hatched region represents the signal
  component.  The dotted, dot-dashed and dashed curves represent,
  respectively, the cumulative background contributions from continuum
  processes, $b\to c$ decays, and charmless $B$ decays.}
\label{hh}
\end{figure}

The CKM angle $\phi_2$ have been measured in decay modes like $B^0\to
\pi\pi, \rho\rho,\rho\pi$~\cite{pipirhorho}.  Addition of new decay
modes allows to improve an accuracy of $\phi_2$ measurement and to
check a consistency of measurements in different final states.  The
decay $B^0\to a_1^\pm(1260)\pi^\mp$ proceeds through $b\to u$
transitions, hence its time-dependent $CP$ violation is also sensitive
to $\phi_2$.  Belle measured the branching fraction for this decay
mode to be ${\cal B}(B^0\to a_1^\pm(1260)\pi^\mp){\cal
  B}(a_1^\pm(1260)\to
\pi^\pm\pi^\pm\pi^\mp)=(14.9\pm1.6\pm2.3)\times10^{-6}$~\cite{a1ppim_belle},
while BaBar has updated their previous measurements now with $CP$
violation study: ${\cal A}_{CP}=-0.07\pm0.07\pm0.02$ and ${\cal
  S}=+0.37\pm0.21\pm0.07$~\cite{a1ppim_babar}.  The angle $\phi_2$ was
measured to be $\phi_2^{\rm eff}=78.6^o\pm7.3^o$.  The result is
presented in Fig.~\ref{hh}(a-c).

Belle also performed the search for the decay $B^0\to\rho^0\rho^0$ and
other decay modes with four pions in the final state.  In the absence
of the signals, the upper limits on the branching fraction were
established.  The signal distributions for the $B^0\to\rho^0\rho^0$
are shown in Fig.~\ref{hh}(a,b).  All results are preliminary.

Also a number of the decay modes potentially usable for the $\phi_2$
measurements have been studied by
BaBar~\cite{a1pi_babar,a1k_babar,b1pik_babar}.  All the results of
these studies are summarized in Table~\ref{hh_table}.
\begin{table}
\centering
\caption{Fit results for decays relevant to $\phi_2$ measurements.}
\begin{tabular}{|lccccc|}
\hline
Mode               & Yield  & $\epsilon$ (\%) & ${\cal S} (\sigma)$ & ${\cal B}(\times10^{-6})$ & UL$(\times10^{-6}) @ 90\%$ C.L. \\
\hline
Belle results & & & & & \\
\hline
$\rho^0\rho^0$     &   $24.5^{+23.6+9.7}_{-22.1-9.9}$ & 9.16 & 1.0 & $0.4\pm0.4^{+0.2}_{-0.2}$ & $<1.0$ \\
$\rho^0\pi^+\pi^-$ & $112.5^{+67.4+51.5}_{-65.6-53.7}$ & 2.90 & 1.3 & $5.9^{+3.5+2.7}_{-3.4-2.8}$ & $<11.9$ \\
$4\pi^\pm$         & $161.2^{+61.2+26.0}_{-59.4-28.5}$ & 1.98 & 2.5 & $12.4^{+4.7+2.0}_{-4.6-2.2}$ & $<19.0$ \\
$\rho^0f^0$        &   $-11.8^{+14.5+4.9}_{-12.9-3.6}$ & 5.10 & $-$ & $-$ & $<0.6$ \\
$f^0f^0$           &      $-7.7^{+4.7+3.0}_{-3.5-2.9}$ & 2.75 & $-$ & $-$ & $<0.4$ \\
$f^0\pi^+\pi^-$    &   $6.3^{+37.0+18.0}_{-34.7-18.1}$ & 1.55 & $-$ & $0.6^{+3.6}_{-3.4}\pm1.8$ & $<7.3$ \\
\hline
BaBar results & & & & & \\
\hline
$b_1^0\pi^+$     &  $178^{+39}_{-37}$ & 6.78 & 4.0 &  $6.7\pm1.7\pm1.0$ & \\
$b_1^0  K^+$     &  $219^{+38}_{-36}$ & 6.73 & 5.3 &  $9.1\pm1.7\pm1.0$ & \\
$b_1^\mp\pi^\pm$ &  $387^{+41}_{-39}$ & 9.54 & 8.9 & $10.9\pm1.2\pm0.9$ & \\
$b_1^-K^+$       &  $267^{+33}_{-32}$ & 9.43 & 6.1 &  $7.4\pm1.0\pm1.0$ & \\
\hline
$a_1^0\pi^+$ & $382\pm79$ &  7.2 & 3.8 & $20.4\pm4.7\pm3.4$ & \\
$a_1^-  K^0$ & $241\pm32$ &  9.6 & 6.2 & $34.9\pm5.0\pm4.4$ & \\
$a_1^+\pi^0$ & $459\pm78$ & 12.5 & 4.2 & $26.4\pm5.4\pm4.2$ & \\
$a_1^-  K^+$ & $272\pm44$ &  7.9 & 5.1 & $16.3\pm2.9\pm2.3$ & \\
\hline
\end{tabular}
\label{hh_table}
\end{table}

\section{$CP$-violation in $\Upsilon(4S)$ decays}

In the decay $\Upsilon(4S)\to B^0\bar B^0\to f_1f_2$, where $f_1$ and
$f_2$ are $CP$ eigenstates, the $CP$ eigenvalue of the final state
$f_1f_2$ is $\xi=-\xi_1\xi_2$.  Here the minus sign corresponds to odd
parity from the angular momentum between $f_1$ and $f_2$.  If $f_1$
and $f_2$ have the same $CP$ eigenvalue, i.e. $(\xi_1,\xi_2)=(+1,+1)$
or $(-1,-1)$, $\xi$ is equal to $-1$.  Such decays, for example
$(f_1,f_2)=(J/\psi K_S^0,J/\psi K_S^0)$, violate $CP$ conservation
since the $\Upsilon(4S)$ meson has $J^{PC}=1^{--}$ and thus has
$\xi_{\Upsilon(4S)}=+1$.  The branching fraction within the SM is
suppressed by the factor
\[
F\approx\frac{x^2}{1+x^2}(2\sin2\phi_1)^2=0.68\pm0.05,
\]
where $x=\Delta m_d/\Gamma=0.776\pm0.008$~\cite{PDG}.

\begin{figure}[htb]
\centering
\includegraphics[width=.33\textwidth]{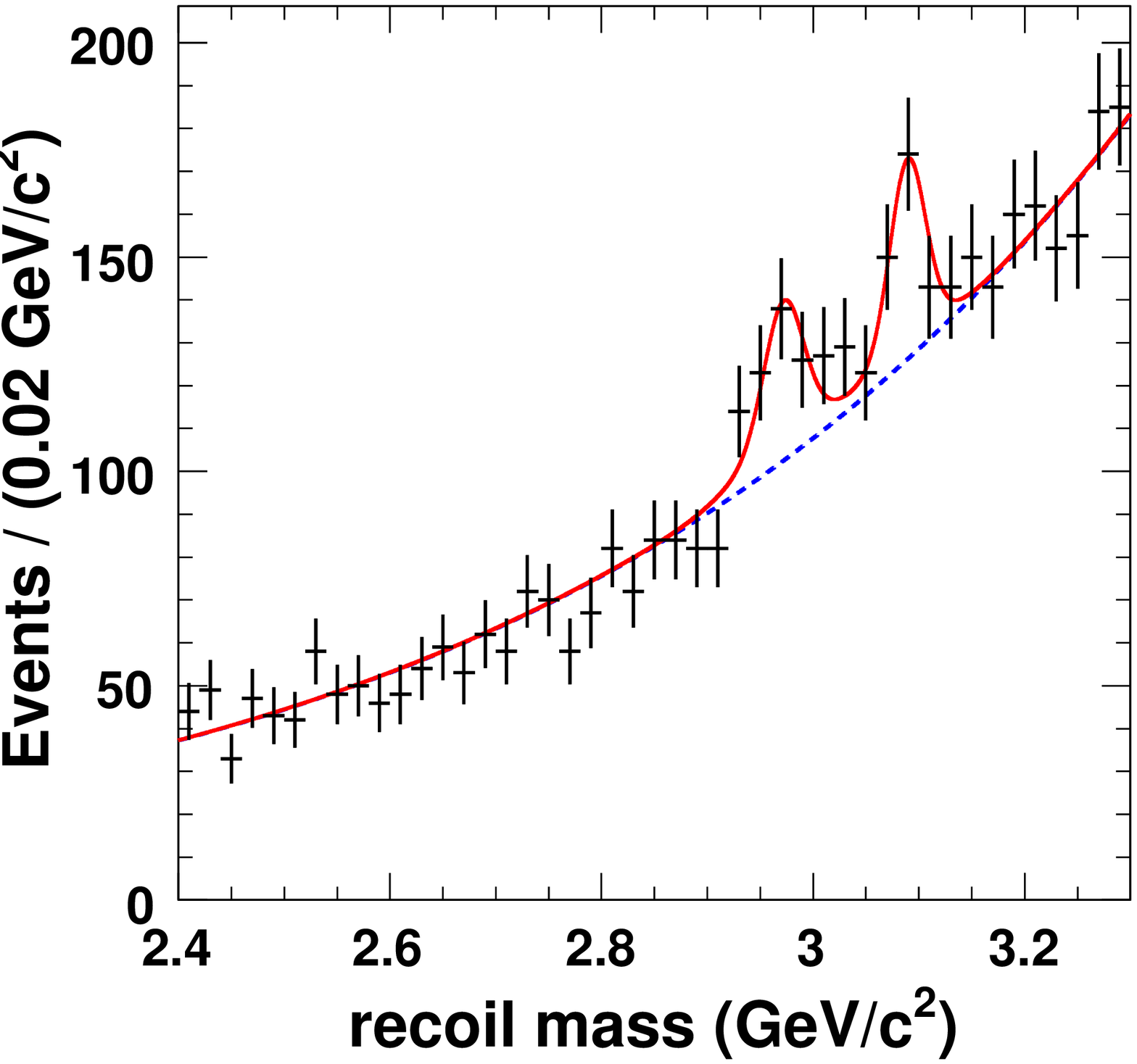}
\put(-120,120){\bf (a)}
\includegraphics[width=.33\textwidth]{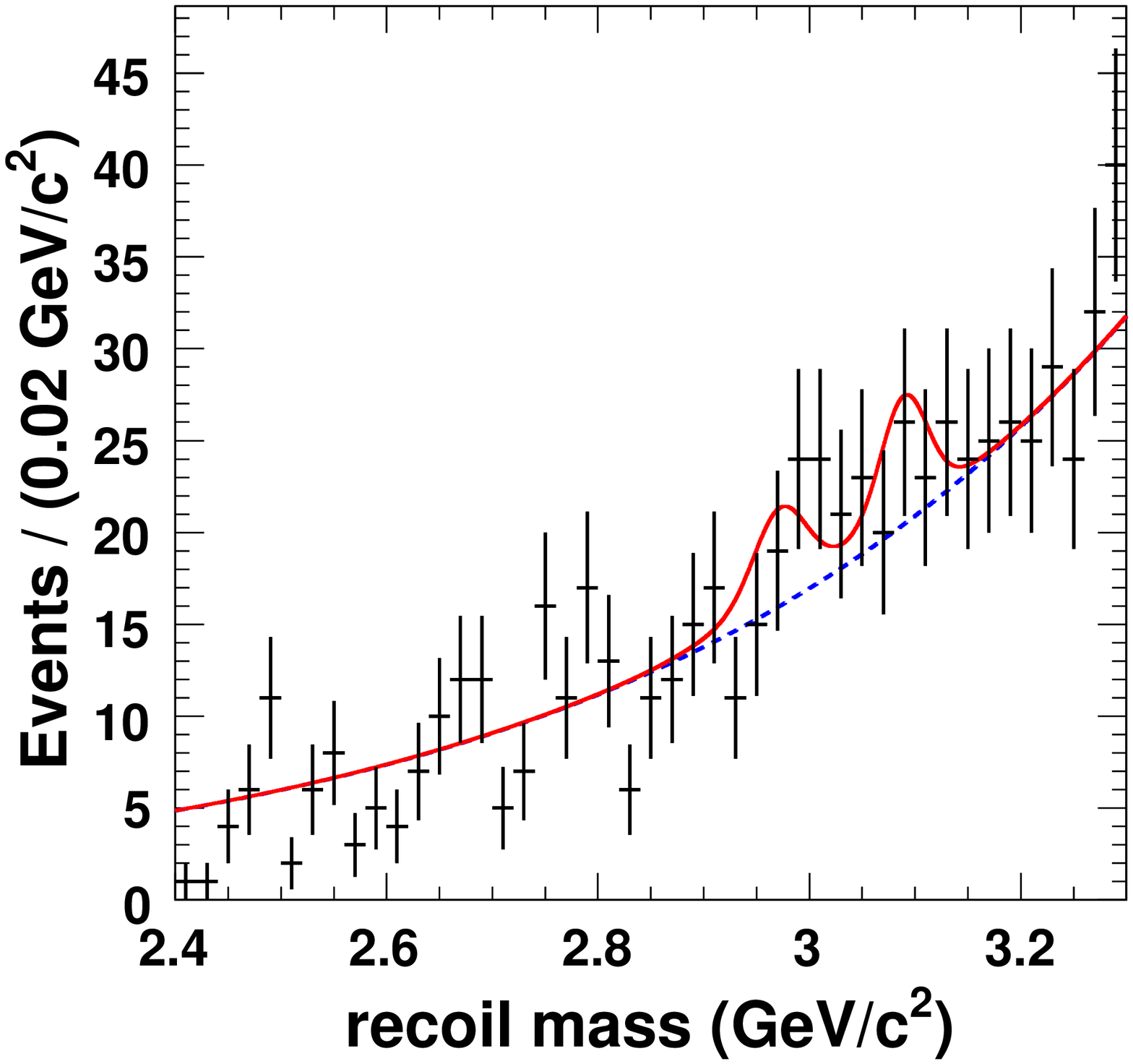}
\put(-120,120){\bf (b)}
\includegraphics[width=.33\textwidth]{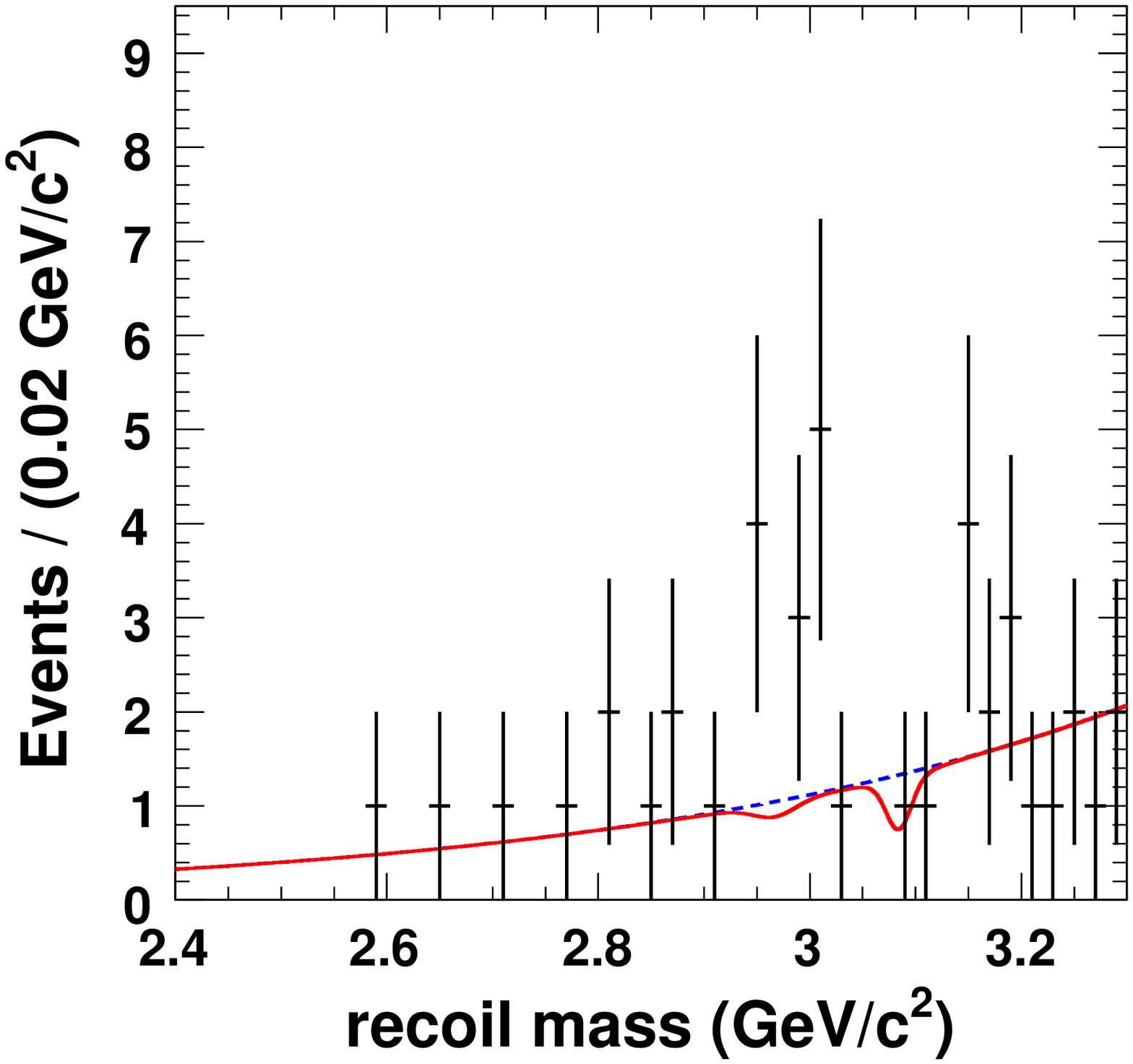}
\put(-120,120){\bf (c)}
\caption{Recoil mass distributions for samples reconstructed as
  $\Upsilon(4S)\to (B^+, (J/\psi,\eta_c)^{\rm tag}K^-)$ (a), $(B^0\to
  D^{(*)-}h^+, (J/\psi,\eta_c)^{\rm tag}K_S^0)$ (b) and $(J/\psi
  K_S^0, (J/\psi,\eta_c)^{\rm tag}K_S^0)$ (c).  The solid lines show
  the fits to signal plus background distributions while the dashed
  lines show the background distributions.}
\label{ups4s_belle}
\end{figure}

This decay was studied by Belle.  Due to a small branching fractions
to the final state and low reconstruction efficiencies the expected
yield is very small, $0.04$ events.  In order to increase the signal
yield, a partial reconstruction technique was used~\cite{ups4s_belle}.
One $B^0$ was fully reconstructed, while only $K_S^0$ was
reconstructed from another one.  The signal was searched in the recoil
mass distribution to the reconstructed particles where, in principle,
signals from $\eta_c,J/\psi,\chi_{c1}$, or $\psi(2S)$ can be seen.
The method was checked using charged $B$ decay control samples,
$\Upsilon(4S)\to B^+B^-\to(f_{B^+},J/\psi^{\rm tag}K^-$ and
$\eta_c^{\rm tag}K^-)$, where $f_{B^+}$ stands for $J/\psi K^+$ and
$\bar D^0\pi^+$.  Also neutral $B$ decays were examined in the decay
$\Upsilon(4S)\to B^0\bar B^0\to(f_{B^0},J/\psi^{\rm tag}K_S^0$ and
$\eta_c^{\rm tag}K_S^0)$ with $f_{B^0}=B^0\to D^{(*)-}h^+$.  The fit
yields $206\pm57$ for charged $B$ and $35\pm16$ for neutral $B$ signal
events, which is in good agreement with the MC expectation
(Fig.~\ref{ups4s_belle}(a,b)).  The results of the final fit are shown
in Fig.~\ref{ups4s_belle}(c).  The extracted signal yield,
$-1.5^{+3.6}_{-2.8}$ events, is consistent with zero as well as with
the SM prediction (1.7 events).  An upper limit for the branching
fraction was obtained ${\cal B}(\Upsilon(4S)\to B^0\bar B^0\to J/\psi
K_S^0, (J/\psi,\eta_c)K_S^0)<4\times10^{-7}$ at the $90\%$ confidence
level, where the SM prediction is $1.4\times10^{-7}$.  This
corresponds to $F<2$ at the $90\%$ confidence level.

\section{Summary}

The $CP$ violating parameters have been measured in various decay
modes.  Most of the measurements are in a good agreement with the SM
expectations.  Although a room for New Physics becomes smaller and
smaller, there is still some sign that it can be found in $b\to s$
transitions.  More statistics is necessary to test these
possibilities.


\begin{thebibliography}{99}
\bibitem{cc} The inclusion of charge conjugate modes is implied
  throughout this Letter.
\end{thebibliography}
\end{document}